\documentclass[twocolumn,english,aps,prb,floatfix,preprintnumbers,
showpacs,amsfonts,amssymb,superscriptaddress]{revtex4}
\usepackage[T1]{fontenc}
\usepackage[latin1]{inputenc}
\usepackage{amsmath}
\usepackage{graphicx}
\usepackage{amssymb}

\usepackage{amscd}
\usepackage{bm}
 
\usepackage{babel}
\makeatother

\newcommand{\be}{\begin{equation}}
\newcommand{\ee}{\end{equation}}
\newcommand{\bt} { \begin{tabular} }
\newcommand{\et}{ \end{tabular} }
\newcommand{\bc} { \begin{center} }
\newcommand{\ec}{ \end{center} }

\newcommand{\f}{ \frac }

\newcommand{\la}{\label }
\newcommand{\bfi}{\begin{figure} }
\newcommand{\efi}{\end{figure} }

\newcommand{\btb} { \begin{table} }
\newcommand{\etb}{ \end{table} }

\begin{document} 
%
\title{Phase Diagram and  Critical Behavior of the
Spin-1 Baxter-Wu Model with a Crystal Field}

\author{M. L. M. Costa}
\affiliation{Departamento de F\'\i sica, Instituto
de Ci\^encias Exatas, Universidade Federal de Minas Gerais, C.P. 702, 
Belo Horizonte, MG 30123-970, Brazil}
\author{J. C. Xavier}
\affiliation{
Instituto de F\'{\i}sica Gleb Wataghin, Unicamp, C.P. 6165, Campinas,
SP 13083-970, Brazil
}
\author{J. A. Plascak}
\affiliation{Departamento de F\'\i sica, Instituto
de Ci\^encias Exatas, Universidade Federal de Minas Gerais, C.P. 702, 
Belo Horizonte, MG 30123-970, Brazil}
\date{\today }
\vspace{0.2cm}
\begin{abstract}
The phase diagram and critical behavior of the spin-1 Baxter-Wu 
model with a crystal field in two dimensions  is  explored by 
renormalization group, conventional finite-size scaling and  conformal 
invariance techniques. 
We found that the phase diagram of this model is qualitatively
the same as that of the dilute 4-states Potts model, presenting a 
multi-critical point for a finite value of the crystal field,
in disagreement 
with previous work based on finite-size calculations. 
However,  our results indicate
that the critical exponents vary continuously  along 
the second-order transition line, differently from the expected 
behavior of the dilute 4-states Potts model.
\end{abstract}
\pacs{64.60Kw, 64.60Cn, 64.60Fr} 

\maketitle
\section{Introduction}
The  Ising model \cite{ons} was the first non-trivial
model exactly solvable in two  dimensions  which exhibits spontaneous 
symmetry breaking. It  became the most popular 
ferromagnetic model in statistical mechanics and even today is the
 object of several studies
in other contexts like random systems.\cite{priv} 
The dynamics of the Ising model is described by the Hamiltonian
\be 
H_I=-J\sum_{<i,j>}s_is_j,
\la{his}
\ee
where the sum is over all nearest neighbors and the classical spin 
variables $s_i=\pm1$ are attached at each site $i$ of the
lattice. In middle 60's,  
Blume and Capel \cite{s1mf} proposed an extension
of the Hamiltonian (\ref{his}) to study first-order magnetic phase 
transitions. 
Their Hamiltonian is given by
\be 
H_{BC}=-J\sum_{<i,j>}s_is_j+\Delta\sum_{i}s_i^2,
\la{hbc}
\ee
where $\Delta$ plays the role of a crystal field and in this 
case the variables $s_i$ are classical spin-1 variables taking the  values  
$s_i=-1,0,1$. 

It is well established that for dimensions $d\ge 2$ the 
Blume-Capel model (Eq. (\ref{hbc})) 
presents a phase diagram with ordered ferromagnetic and disordered
paramagnetic phases separated by a transition line which changes from a
second-order character (Ising type) to a first-order one at a tricritical
point (see  Ref. \onlinecite{new} and references there in).
More specifically, in two dimensions, the machinery coming from
conformal invariance \cite{ci1,ci2} indicates that at this multi-critical
point the long-range fluctuations are governed by a conformal field theory 
with central charge $c=7/10$.\cite{new,s1c.7,drugo} 
In this case, all the critical exponents and the whole operator content of 
the model were obtained \cite{drugo} (see also Ref. \onlinecite{new}). 
The generalization to higher spin $S$ of this model
has also been studied.\cite{new,srg,s1.5mk,s1.5mf,plandau0} 
In particular,  results of mean field  theory,\cite{s1.5mf} 
conformal invariance,\cite{new} and 
Monte Carlo simulations \cite{plandau0} predict different phase diagrams  
for integer or half-odd-integer  spins, in contradiction with 
results based on real space renormalization groups.\cite{srg,s1.5mk}
Recently, the universality at a double critical endpoint in the
two-dimensional spin-3/2 Blume-Capel model has been analyzed and it
was shown that it belongs indeed to the same universality class as the 
critical line.\cite{plandau}

Another simple model exactly solvable in two dimensions exhibiting 
spontaneous symmetry breaking  is the  Baxter-Wu model.\cite{bw,ax1,ax2}  
This model is defined  on a triangular lattice  by the three-spin 
interaction Hamiltonian
\be
H_{BW}=-J\sum_{<ijk>}s_is_js_k ,
\la{hbw}
\ee
where the sum extends  over all  elementary triangles of the lattice   
and $s_i=\pm 1$ are Ising variables located at the sites.
This model is  self-dual \cite{merlini}
with the same critical temperature as that of  the Ising model on a 
square lattice. The critical behavior  of the Baxter-Wu model 
is  governed by a conformal field theory  with central charge $c=1$, 
\cite{ax1,ax2}
 and its leading exponents \cite{bw,ax1,ax2} $\alpha =2/3$, $\nu=2/3$ and  
$\eta=1/4$ are  the same as those of the 4-states Potts model. 
\cite{dwbaxter}

In analogy with the Blume-Capel model, in this paper we consider the  
Baxter-Wu model in the presence of a crystal field. 
The Hamiltonian of the spin-1 Baxter-Wu model 
with a crystal field  is  given by 
\be
H=-J\sum_{<ijk>}s_is_js_k +\Delta\sum_{i}s_i^2,
\la{hbcbw}
\ee
where  the classical spin variables $s_i$, defined in a triangular lattice,
take the values $s_i=-1,0,1$. Note that when
$\Delta\rightarrow-\infty$ only the configurations with $s_i=\pm1$ 
are allowed, and we recover the pure Baxter-Wu model.

Since we have in the Baxter-Wu model with a crystal field the same kind of  
competition between the ordered  ($\langle s\rangle \ne 0$) and the 
disordered phases ($\langle s\rangle= 0$) (which  is mediated by the crystal
field) as in the Blume-Capel 
model, we may expect for both models  a similar phase diagram, but with 
different critical behavior. This kind of competition  
also appears in the dilute $q$-states Potts model.\cite{nis}
It is well known that the Baxter-Wu model and the 4-states Potts model
have the same critical exponents  (see for example Ref. 14). Since the
dilution in the 4-states Potts model has the same effect as the
crystal field in the Baxter-Wu model, we may expect  the critical behavior 
of both models to be the same. Some previous calculations of the phase 
diagram of both models have been reported in the literature. Nienhuis 
{\it et al.\/},\cite{nis} based on a renormalization-group study, 
indicate that 
for the dilute 4-states Potts model  the phase diagram is similar as 
that of the Blume-Capel model, i.e., there is a transition
line which changes from a second-order character to a first-order one at 
a multi-critical point. In this case, however, the critical behavior is
 governed
by only one fixed point, giving along the second-order line the same
exponents as that of the pure 4-states Potts model.  
On the other hand, Kinzel {\it et al.\/},\cite{kda}
using finite-size methods, conjectured a different kind of phase diagram for 
the Baxter-Wu model in the presence of a  crystal field. 
These authors interpreted the
changes of the estimated thermal exponents $y_t$ along the transition line
as a signal that a second-order transition  should happen only for
$\Delta\rightarrow-\infty$ (the pure Baxter-Wu model).

A careful study of the pure Baxter-Wu model has also been done by 
exploring its Bethe-ansatz solution.\cite{bw,ax1,ax2} Using  
the consequences of conformal invariance, it has been shown \cite{ax1,ax2} 
that in the absence of dilution ($\Delta\rightarrow-\infty$) not only the
leading critical exponents of the Baxter-Wu model and the 4-states Potts 
model are identical, but the whole operator content of the models coincides
as well. Moreover, the masses of the field theory describing the thermal 
and magnetic perturbations of both models are also identical.\cite{ax2} 
Since the reported effects of the dilution in both systems are different,  
we decided in this paper to study the effect of dilution (or crystal field) 
in the  Baxter-Wu model in two dimensions using the machinery of 
conformal invariance and renormalization group techniques.

This paper is organized as follows. In the
next section we present the phase diagram obtained through the mean
field renormalization group approach. In section III we present the
transfer matrix of the model,
the relations used in our finite-size studies, as well as the
results for the phase diagram. We then close the paper in
section IV with a summary and conclusions.
\section{Mean field renormalization group}
The mean field renormalization group (MFRG)\cite{indekeu1,indekeu2} 
is a powerful phenomenological
approach which can provide quite good results  in the general 
study of critical phenomena.\cite{plarev}  
It is based on the comparison of the order
parameter for different finite lattices in the presence of symmetry breaking 
boundary conditions. For a finite cluster of $N$ spins, and
considering the parameters of the Hamiltonian (\ref{hbcbw}), one first computes
the magnetization per spin $m_N(K,D,b)$, where $K=\beta J$, 
$D = \beta \Delta$ and $b$ is the boundary field, with $\beta =1/k_BT$ and
$k_B$ the Boltzmann constant. As the boundary field is assumed to be very 
small one has
\be
m_N(K,D,b)=f_N(K,D)b~.
\label{mag}
\ee

In its simplest version,  the MFRG considers two different clusters of 
$N^\prime<N$ spins and assumes that the magnetizations scale as
\be
m_{N^\prime}(K^\prime,D^\prime,b^\prime)=\ell^{d-y_H}
m_{N}(K,D,b),
\label{mscale}
\ee
where $\ell=(N/N^\prime)^{1/d}$ is the scaling factor, $d$ is the dimension 
of the lattice and $y_H$ the magnetic exponent. With the same relation for
the boundary fields, namely
\be
b^\prime=\ell^{d-y_H}b,
\label{bsclae}
\ee
and taking into account expansion (\ref{mag}) one has
\be
f_N(K,D)=f_{N^\prime}(K^\prime,D^\prime),
\label{smfrg}
\ee
which is independent of any scaling factor and is viewed as a renormalization
recursion relation from which one gets fixed point solutions $K^\prime=K=
K_c$ and estimates of correlation length critical exponents $\nu$ in the 
subset $D^\prime=D$.

Another version of the approach considers three different clusters of 
$N^{\prime\prime}$, $N^\prime$ and $N$ spins (in increasing order)
together with the scaling (\ref{mscale}) in such a way that one gets
\be
f_{N^\prime}(K^\prime,D^\prime)b^\prime=\ell_1^{d-y_H}f_N(K,D)b,
\label{3smag1}
\ee
\be
f_{N^{\prime\prime}}(K^{\prime\prime},D^{\prime\prime})b^{\prime\prime}=
\ell_2^{d-y_H}f_{N^\prime}(K^\prime,D^\prime)b^\prime,
\label{3smag2}
\ee
where the scaling factors are $\ell_1=(N/N^\prime)^{1/d}$ and 
$\ell_2=(N^\prime/N^{\prime\prime})^{1/d}$. Imposing now that the boundary 
fields scale not as bulk magnetizations above, but as the surface field we
obtain
\be
b^\prime=\ell_1^{y_{hs}}b,~~~~~~b^{\prime\prime}=\ell_2^{y_{hs}}b^\prime ,
\label{3sb}
\ee
where $y_{hs}$ is the surface critical exponent. 
Eqs. (\ref{3smag1})-(\ref{3sb})
are now the renormalization recursion relation between the interaction
parameters of the system. The exponent $d-y_h-y_{{hs}}$ is then determined
self-consistently by imposing further that Eqs. (\ref{3smag1}) and 
(\ref{3smag2})  possess the same fixed point 
$K^{\prime\prime}=K^{\prime}=K=K_c$
for an invariant subset $D^{\prime\prime}=D^{\prime}=D$
(for further details and a comparison between these two approaches
see Refs. \onlinecite{indekeu2} and \onlinecite{plarev}). 
This version of the method is 
referred as surface bulk MFRG (SBMFRG).

Before applying the MFRG  
to the spin-1 Hamiltonian,  defined by Eq. 
(\ref{hbcbw}), it is worthwhile to measure the efficiency of the method
by first treating the pure Baxter-Wu model (Eq. \ref{hbw}), 
where exact results are
available. Since the model exhibits three different ferrimagnetic
phases at low temperatures,\cite{comm} it is supposed, at first sight, 
that the size of the finite blocks must be such that they will suitably 
accommodate them in an equivalent way throughout the lattice.
These would dramatically restrict the size of the clusters. However,
we note that by computing the sub-lattice magnetizations
$m_A$, $m_B$ and $m_C$, by taking three different boundary fields
$b_A$, $b_B$ and $b_C$, we obtain the same equations as by considering
a homogeneous cluster where $m_A=m_B=m_C$ and $b_A=b_B=b_C$. This is not
surprising, bearing in  mind that the ferromagnetic phase is also
coexisting with the other three ferrimagnetic ones at low temperatures.
This means that, for practical purposes, we can consider only the
ferromagnetic arrangement, allowing us to take blocks of any number of spins.
This leads a huge simplification in the numerical acquisition
of the functions $f_N(K)$.
For this particular system, the best clusters are those which preserve
the symmetry of the original lattice. They are made of symmetric triangles 
of $N=6, 10, 15, 21, 28$ spins. Although for $N=6$ we can obtain analytical
expressions for $m_6(K,b)$ and $f_6(K)$, for $N\ge 10$ all the quantities have
to be computed numerically. In Table \ref{tab1a}, we present the critical 
temperature and critical exponent obtained according to the usual MFRG and 
also from the SBMFRG. It is also possible  through 
the MFRG  get an extrapolation. \cite{plarev} Note that, for the present
case, the extrapolated value is not so close to the exact result. This is, in fact,
%
%
\begin{table}[here]
\caption{Results for the pure Baxter-Wu model according to the
MFRG and SBMFRG approaches.}
\bc
\begin{tabular}{||c|c|c||} \hline
$N^\prime-N$ &  $K_{C}$ & $\nu$    \\
\hline \hline
             &   MFRG   &          \\
\hline	     
6-10         &  0.28667 & 2.78203  \\
10-15        &  0.36565 & 1.89869  \\
15-21        &  0.39335 & 1.55875  \\
21-28        &  0.41077 & 1.37178  \\
\hline
extrapolated &  0.43768 & 0.69995  \\
exact        &  0.44069 & 2/3      \\
\hline
             &   SBMFRG &          \\
\hline	     
6-10-15      &  0.58138 & 1.06875  \\
10-15-21     &  0.48090 & 1.08061  \\
15-21-28     &  0.47832 & 0.99573  \\
 \hline
\end{tabular}
\label{tab1a}
\ec
\end{table}
%
%
%
expected since the MFRG does not reproduce the exact value as the size
of the lattices go to infinity.\cite{indekeu2} 
Although the SBMFRG does,\cite{indekeu2}
the present clusters are still too small for us to get a reasonable
extrapolation. However, as can be seen from Table \ref{tab1a}, the temperatures
are getting closer to the exact value as the size of the clusters
increase. Note that, while the MFRG approaches the expected result 
from below, the SBMFRG does it from above. A common feature of the present renormalization 
group is still a worse estimate of the critical exponent when compared to 
the critical temperature.
Nevertheless, in general, a reasonable picture of the criticality of the 
system is achieved from the approach.

\begin{figure}[ht]
\includegraphics[clip,angle=0,width=8.3cm]{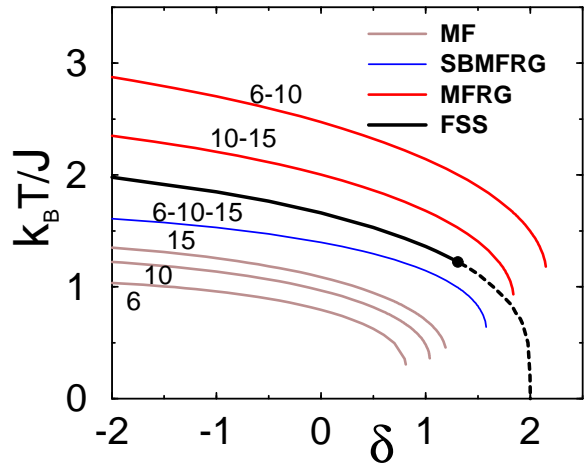}
\caption{\label{phas-diag} Phase diagram in the $\delta$ versus $kT/J$
plane for the spin-1 
Baxter-Wu model in the presence of a crystal field. 
Continuum lines are  second-order phase
transitions and the dashed line is a first-order transition (see next section). 
The dot represent the multi-critical point. The results are according to the MFRG
with two cluster,  SBMFRG with three clusters,
FSS (see next section) and MF approximation  with one cluster. The sizes of
the clusters used are also presented. 
}
\end{figure}

We now proceed to the study of the spin-1 model with crystal field anisotropy.
The same symmetry arguments also apply for this model. However, due to
computer time, we were here limited to block sizes  $N=6, 10, 15$. 
The phase diagram in the $\delta =\Delta/J$ versus $kT/J$ plane is
depicted in Fig. \ref{phas-diag} according to both procedures, as well
%
as from the finite-size scaling (FSS) procedure of the next section,
and the usual mean field (MF) approximation.
The latter approach, is obtained by assuming
$m_N=b$ in Eq. (\ref{mag}), resulting in $f_N=1$. 
Except for the FSS, all the lines
terminate at some point which is identified as the multi-critical
point. The first-order transition lines are not possible to be obtained
neither from the MFRG nor from the SBMFRG. Comparing with the FSS
result, discussed in the next section, we can see that the MFRG overestimates
the critical temperature while the surface bulk version underestimates it.
This is just what happens
in the pure case, as shown in Table \ref{tab1a},
when compared to the exact critical value. 
( nao entendi o q vc quer dizer...
 In all cases, the critical temperature of the spin-1/2 model is obtained in the limit
$\delta \rightarrow -\infty$. pode tirar esta frase? Ja' nao foi dita na introducao!) 
The estimate of the multi-critical point
is given in Table \ref{tab1b}. 
The SBMFRG and MFRG give critical exponents that vary along
the second-order critical line. However, this fact, may be just an artifact
of  these aproaches, since there is only one  renormalization recursion
relation. \cite{plarev} In the next section, we analize in the
context of confomal invariance the possibility of the critical exponents
vary along the second-order line. 


%
\begin{table}[here]
\caption{Position of the multi-critical point 
for the curves shown in Fig. \ref{phas-diag}.}
\bc
\begin{tabular}{||c|c|c||}
 \hline
$N^\prime-N$ &$k_BT_{t}/J$ & $\delta_t$\\
\hline \hline
             &   MFRG   &         \\
\hline	     
6-10         &  1.1816  & 2.1523  \\
10-15        &  0.9330  & 1.8462  \\
\hline
             &   SBMFRG &         \\
\hline	     
6-10-15      &  0.6408  & 1.5835  \\
\hline
             &   MF     &         \\
\hline	
6            &  0.3133  & 0.8141  \\
10           &  0.3539  & 1.0328  \\
15           &  0.4513  & 1.1902  \\
\hline
             &   FSS     &         \\
\hline	     
3-6-9          &  1.2225  & 1.3089  \\
 \hline
\end{tabular}
\label{tab1b}
\ec
\end{table}
%

%
%
%
\section{Finite-Size Scaling and Conformal Invariance } 
The row-to-row transfer matrix $\hat T$ of the Hamiltonian (\ref{hbcbw}) in a
triangular  lattice, with horizontal width $N$, has dimension $3^N\times
3^N$. Its coefficients $\langle s_1^{\prime },...,s_N^{\prime }|\hat
T|s_1,...,s_N\rangle$ are the Boltzmann weights generated by the spin
configurations $\{s_1,...,s_N\}$ and $\{s_1^{\prime },...,s_N^{\prime }\}$
of adjacent rows. If we consider periodic boundary condition in the horizontal
direction, the transfer matrix can be written as
\begin{eqnarray}
\langle s_1,...,s_N|\hat T|s_1',...,s_N'\rangle= \nonumber
\end{eqnarray}
\begin{eqnarray}
\prod_{j=1}^{N}
\exp
\left[
t^{-1} s_{i+1}s_i'(s_i+s_{i+1}')-\delta t^{-1} s_i^2 
\right] ,
\la{tm}
\end{eqnarray}
with $t=k_BT/J$ ($k_B$ is the Boltzmann constant) and $\delta=D/J$.

The finite-size behavior of the eigenvalues of 
$\hat T$ $(\Lambda_0(N)>\Lambda_1(N),...)$
can be used to determine the critical line and  the 
critical exponents.\cite{ci1,ci2,fss} 
The critical line $(t_c(\delta))$ is evaluated by
 extrapolating to the bulk limit ($N\rightarrow \infty $)
  the sequences $t_c(\delta,N)$ obtained by solving
\be
G_N(t_c)N=G_{N+3}(t_c )(N+3),~~N=3,6,...
\la{fss}
\ee
where $G_N(t_c)$ is the mass gap of $H=-\ln\hat T$ and is given by 
$$
G_N(t_c)=
\ln
\left(\f{\Lambda_0(N)}{\Lambda_1(N)}\right).
$$
The multi-critical points are obtained using a heuristic method, 
which has already been proved to be effective.\cite{new,tripoint}
In this case we have to solve simultaneously  (\ref{fss}) for three 
different lattice sizes
\begin{eqnarray}
G_N(t_c)N=G_{N+3}(t_c )(N+3)= \nonumber
\end{eqnarray}
\begin{eqnarray}
G_{N+6}(t_c)(N+6),~~N=3,6,...
\la{tripoint}
\end{eqnarray}
In Eqs. (\ref{fss}) and (\ref{tripoint}) we restricted the possible 
finite strip widths to multiples of 3  in order to preserve
the invariance  of  the Hamiltonian (\ref{hbcbw}) under the  reversal of all 
spins on any two sub-lattices.

As usual, we expect the model to be conformally invariant in 
the region of continuous phase transition. This invariance allows 
us to infer the critical properties from the finite-size corrections
of the eigenspectrum at $t_c$.\cite{ci1,ci2} 
 The conformal anomaly $c$, which 
labels the universality class of critical behavior, can be
calculated from the large-$N$ behavior 
of the ground-state energy of $H=-\ln\hat T$\cite{ci2}
\be
\frac{\ln\Lambda_0(N)}{N}= \epsilon_{\infty} + \frac{\pi c v_{s}}{6N^{2}} 
+o(N^{-2}),
\la{ano}
\ee
where $\epsilon _\infty $ is the the ground-state energy 
per site in the bulk limit and  $v_s=\sqrt3/2$ is the sound velocity. 
The scaling dimensions of operators governing the critical
fluctuations (related to the critical exponents) are evaluated from the
finite-$N$ corrections of the excited states. For each primary operator, with
dimension $x_\phi $, in the operator algebra of the system, there exists an
infinite tower of eigenstates of $H=-\ln\hat T$ whose energy
 $\ln(\Lambda_{m,m^{\prime}}^\phi)$
and momentum $P^{\phi}_{m,m'}$ are given by \cite{ci1}
\begin{eqnarray} 
\f{\ln\Lambda_{m,m'}^{\phi}(N)}{N}&=&\frac{\ln\Lambda_0(N)}{N}
-\frac{ 2\pi v_s }{ N^2 }(x_{\phi}+m+m')  \nonumber \\
P_{m,m'}^{\phi}&=&\f{2\pi}{N}(s_{\phi}+m-m'), \la{dim}
\end{eqnarray}
where $m,m^{\prime }=0,1,\dots$. 

A finite-size estimate for the first-order transition line can be
obtained by the same procedure done as in a recent work.\cite{new} 
At the first-order line of the Baxter-Wu model with a crystal field,
we have the coexistence of five phases, one ordered ferromagnetically,
three ordered ferrimagnetically 
and a disordered one. Consequently, for a given lattice size $N$ we 
calculate the points where the gap corresponding to the fifth eigenvalue 
has a minimum. The extrapolation $N\rightarrow \infty $ 
of these points give us an estimate for the first-order transition line.

In the numerical diagonalization of (\ref{tm}) we used the Lanczos method for
non-Hermitian matrices.\cite{lanc} We also considered the translational
symmetry to block-diagonalize the transfer matrix.

In Fig. \ref{phas-diag} we show the   second-order transition line 
(continuum line), 
obtained by solving Eq. (\ref{fss}) for lattice sizes $N=6$.
 As we can see in this figure the second-order transition
 line  also occur  for finite values of the crystal field 
($\delta\ne-\infty)$, 
differently of the previous results of Kinzel {\it et al.\/},\cite{kda} 
where it was conjectured the appearance of the second-order transition 
 only at $\delta\rightarrow-\infty$. For the pure Baxter-Wu model
 ($\delta\rightarrow -\infty$) we obtained $t_c^{-1}=0.4408842$ for $N=6$ in 
Eq. (\ref{fss}), 
which differs only  $0.04\%$  of the exact value  
$t_c^{-1}=\f{ \ln(\sqrt{2}+1) }{2} 
=0.440686...$ . For this reason, it is a very good approximation to 
consider $t_c(\delta)=t_c^{6,9}(\delta)$.

For the sake of clarity, we present in Table \ref{tab2} the finite-size 
sequences 
obtained by solving Eq. (\ref{fss}) for lattice sizes $N=3$ and $N=6$. 
 As we can see from this table the convergence of $t_c$ are 
better for $\delta>0$, so we expect that the estimates of 
$t_c^{(6,9)}(\delta)$ are
better than the corresponding ones in the region $\delta<0$.
The fast convergence with $N$ indicates that the corrections to finite-size 
are probably given  by  a power law,  like in the pure 
Baxter-Wu model. In this last case the corrections are controlled 
by an operator with dimension  $w=4$.\cite{ax1}

%
\btb[h]\caption{Finite-size data $t_c^{N,N+3}$ given by (\ref{fss})
 for  the critical temperature $t_c$ for some values of $\delta$.}
\bc
\bt{||c|c|c|c|c|c||} \hline
 N &  -10      & -1        &  1       & 1.25     &  1.3       \\ \hline\hline
 3 &  2.246498 &1.843818   &1.358399  & 1.251121 & 1.226940   \\ \hline
 6 &  2.256769 &1.849705   &1.360144  & 1.251529 & 1.227005   \\ \hline 
\et
\ec
\label{tab2}
\etb
%

We  have also solved Eq. (\ref{tripoint}) for $N=3$, which gives us 
an estimate for the multi-critical point. In this case 
we do not have points to extrapolate, but we believe the estimate 
of the multi-critical point, the last point in the continuum line in Fig.
\ref{phas-diag}, 
is not far from the extrapolated one. Our estimate for this point is
$t_t=1.2225$ and $\delta_t=1.3089$.

We determined the first order line minimizing the gap related with the 
fifth eigenvalue, as discussed before. The dashed line  shown in 
Fig. \ref{phas-diag}  
was obtained in this procedure considering $N=9$. As we can see in this 
figure, the first-order transition line finishes at the multi-critical point.

In the critical regions of the phase transition line (continuum curve) the
conformal anomaly and the scaling dimensions can be calculated exploring the
conformal invariance relations (\ref{ano}) and (\ref{dim}). 
From Eq.\ (\ref{ano}) a possible way to extract $c$ is by extrapolating
the sequence
\begin{eqnarray}
c^{N,N+3}=\f{12}{\sqrt{3}\pi}
\left( \f{\ln\Lambda_{1,0}(N+3)}{N+3}-\f{\ln\Lambda_{1,0}(N)}{N}
 \right)\times \nonumber
\end{eqnarray}
\begin{eqnarray}
\left( \f{1}{(N+3)^2}-\f{1}{N^2} \right)^{-1},
\la{anoest}
\end{eqnarray}
calculated at $t_c(\delta)$. In Eq. (\ref{anoest})  
$\Lambda_{n,p}(N)$ means the 
$n$th  eigenvalue of (\ref{tm}) with size $N$ in the sector 
with momentum $p$. Examples of such sequences  for the Baxter-Wu model with 
a crystal field are shown in Table \ref{tab3}. 
The extrapolated values of $c^{\infty}$ can be obtained from the
$N$-large behavior of $c^{N,N+3}(\delta)$, given by
\begin{eqnarray}
c^{N,N+3}(\delta)=c^{\infty}-a\left(N^{-w}+(N+3)^{-w}\right)
 \times \nonumber
\end{eqnarray}
\begin{eqnarray}
\left(N^{-2}+(N+3)^{-2}\right)^{-1},
\la{cin}
\end{eqnarray}
where $a$ is a constant and $w$ the dominant dimension associated
to the operator governing the finite-size corrections. We determine
 $c^{\infty}$  assuming    Eq. (\ref{cin}) exact for $N=3$ and 
$N=6$ with $w$ given by the pure Baxter-Wu model, 
i.e. $w=4$.\cite{ax1}
We see from Table  \ref{tab3}  that the conformal anomaly is $c=1$
along the critical line, and apparently even at the multi-critical point. 
This scenario is quite different from that of the Blume-Capel model, where
the conformal anomaly changes abruptly from $c=1/2$ at the critical 
line to $c=7/10$ at the tricritical point (see Ref. 5), but its 
qualitatively similar as that of the dilute 4-states Potts model,\cite{nis} 
where along of the  critical line the critical exponents as well the 
conformal anomaly do not change. 
 Since we do not have a precise estimate for the multi-critical point, we
have also calculated $c^{N,N+3}$ for several 
values of $t_c(\delta)$ between  $1.2<t_c(\delta)<1.4$. We
have not seen any abrupt change of the conformal anomaly, like in the
Blume-Capel model.\cite{new}

%
\btb[ht]\caption{Finite-size estimates $c^{N,N+3}$, given by 
(\ref{anoest}), for  the conformal anomaly is shown for some values of 
$\delta$.
The  extrapolated results obtained by (\ref{cin}) is also shown.}
\bc
\bt{||c|c|c|c|c|c||} \hline
 N & -10      & -1 &  1   &  1.25     &  1.3089           \\ \hline\hline
 3 &  0.938546&0.941169   &0.954322  & 0.955674 & 0.955047   \\ \hline
 6 &  0.986393&0.986829   &0.986656  & 0.979694 & 0.975435    \\ \hline \hline
$\infty$ & 1.006&1.005 &0.999     & 0.989    & 0.984  \\ \hline
\et
\ec
\label{tab3}
\etb
%

From Eq. (\ref{dim}) the scaling dimensions $x(n,p)$ related to the
$n$th ($n=1,2,\ldots$) energy in the sector with momentum $p$ 
can be obtained by extrapolating the sequence
\be
x^N(n,p)=\f{ N}{\pi\sqrt{3}}
\ln
\left(\f{\Lambda_{1,0}(N)}{\Lambda_{n,p}(N)}\right) .
\la{dimest}
\ee
%
%
%
\btb[ht]\caption{Finite-size scaling dimensions $x^N(1,0)$ and $x^N(2,0)$ given by 
(\ref{dimest})  for some values of $\delta$.
}
\bc
\bt{||l|l||l|l|l||} \hline\hline
          &                   & N=9    &  N=6   &  N=3      \\ \cline{3-5} 
          & $\delta$=-10      & 0.1236 & 0.1236 & 0.1195       \\ \cline{3-5}
          & $\delta$=-1       & 0.1223 & 0.1223 & 0.1184        \\ \cline{3-5}
$x^N(1,0)$& $\delta$=1        & 0.1113 & 0.1113 & 0.1093        \\ \cline{3-5}
          & $\delta$=1.25     & 0.1048 & 0.1048 & 0.1043         \\ \cline{3-5}
          & $\delta$=1.3089   & 0.1026 & 0.1026 & 0.1026         \\ \hline\hline
          & $\delta$=-10      & 0.5059 & 0.5147 & 0.6057         \\ \cline{3-5}
          & $\delta$=-1       & 0.4897 & 0.4976 & 0.5756        \\ \cline{3-5}
$x^N(2,0)$& $\delta$=1        & 0.3644 & 0.3762 & 0.4163       \\ \cline{3-5}
          & $\delta$=1.25     & 0.3023 & 0.3207 & 0.3613       \\ \cline{3-5}
          & $\delta$=1.3089   & 0.2829 & 0.3037 & 0.3456    \\ \hline\hline
\et
\ec
\label{tab4}
\etb
In Table \ref{tab4} we show the dimensions $x^N(1,0)$ and $x^N(2,0)$
for the Baxter-Wu model with a crystal field. Note that $x^9(1,0)=x^6(1,0)$ due
the Eq. \ref{fss} and the fact we choose   $t_c(\delta)=t_c^{6,9}(\delta)$. In the
tricritical point the three entries $x^N(1,0)$ are same due Eq. \ref{tripoint}.
For $\delta\rightarrow -\infty$ (the pure
 Baxter-Wu model) the scaling dimensions  $x^9(1,0)$ and $x^9(2,0)$
differ only $1 \% $ from the leading dimensions
 $x^9(1,0)=\f{1}{8}$ and $x^9(2,0)=\f{1}{2}$
of the pure Baxter-Wu model. As the estimate of the critical line is
better in the region $\delta>0$ and the eigenvalues converge faster with
the size of the lattice in this region, we believe that our estimates for the
scaling dimensions are better in the region $\delta>0$ than the 
corresponding ones for $\delta<0$, i.e., the estimates $x^9(1,0)$ and 
$x^9(2,0)$
must differ by less than  $1 \% $ from the
extrapolated values for $\delta>0$. Note that
 when we increase the crystal field $\delta$ these values change continuously
 up to   $x^9(1,0)\sim 0.10$ and 
$x^9(2,0)\sim 0.28$ at the multi-critical point. This scenario
is quite different from that of the Blume-Capel model, 
which is not a surprise since both models are in different universality
classes of critical behavior. However it is also distinct from
 the scenario of the dilute 4-states Potts
model, where the scaling dimensions along the critical line are 
believed to be the same as that of the pure 4-states Potts model.
 If the dilution had the same role in both models,
we should expect for the dilute 4-states Potts a continuous
line of fixed point, however this was not found.\cite{nis}
\section{Summary and Conclusion}
In this paper we have  calculated the phase  diagram and critical 
properties of the spin-1 Baxter-Wu model in the presence of a crystal 
field. Our results, based on 
renormalization group, finite-size scaling and conformal 
invariance, show a second-order transition line  separated 
from a first-order transition line by a multi-critical point. This scenario 
is in disagreement with that of a previous paper by Kinzel {\it et al.\/}, 
\cite{kda} 
where the second-order transition line  appears only in the 
limiting case  $\Delta \rightarrow -\infty$. 
The  critical behavior was determined by renormalization group and
conformal invariance. Despite the phenomenological renormalization 
group be not so
conclusive regarding the critical exponents, the conformal invariance
results indicate that along the critical line, and even at the 
multi-critical  point,the conformal anomaly is the same as that of the 
Baxter-Wu model or the  4-states Potts model, i.e., $c=1$, in agreement 
with the scenario expected for the dilute 4-states Potts model. 
However, our results indicate
that the scaling dimensions vary continuously with the crystal field.
This is  an unexpected behavior since the reported results for 
the dilute 4-states Potts model indicate a constancy of the scaling 
dimensions along the phase transition line,\cite{nis} and it is expected 
that both models belong to the same universality class of critical
behavior. This result implies that either contrary to what occurs with 
thermal and magnetic perturbations the effect of dilution is distinct in
the Baxter-Wu and in the 4-states Potts models, or the scenario based in
renormalization group for the dilute 4-states Potts model is wrong. 
It would be interesting to verify our results using different numerical 
techniques, like Monte Carlo methods. In fact, a Monte Carlo simulation
for the case $\delta=0$ has already been done and  different exponents as
those of the pure Baxter-Wu model have been achieved.\cite{moraes} Extensive
Monte Carlo simulations for $\delta \ne 0$ are in progress and will 
be present elsewhere. \cite{moraes2}

\begin{center}
{\bf Acknowledgments}
\end{center}
JCX would like  to  acknowledge profitable discussions 
with M. J. Martins and J. R. G. de Mendon\c{c}a.
He is also indebted to F. C. Alcaraz for  suggestions and discussions in 
part of this work. This work was supported  by 
FAPESP (Grants 00/02802-7 and 01/00719-8) (JCX), FAPEMIG (MLMC and JAP), CNPq 
(JAP) and  CAPES (MLMC).


\begin{thebibliography}{99}
%
%
%
\bibitem{ons}
 L. Onsager, {\it Phys. Rev.\/} {\bf 65}, 117 (1944).
%
%
\bibitem{priv}
 M. P. Nightingale,  in {\it  Finite Size Scaling and
Numerical Simulations in Statistical Systems\/}, edited by V. Privman
(World Scientific, Singapore, 1990). 
%
%
%
\bibitem{s1mf}
 M. Blume, {\it Phys. Rev.\/} {\bf 141}, 517 (1966);
H. W. Capel, {\it Physica\/} {\bf 32}, 966 (1966).
%
%
%
\bibitem{new} J. C. Xavier, F. C. Alcaraz, D. P. Lara , and J. A. Plascak,
 {\it Phys. Rev. B\/} {\bf 57}, 11575 (1998).
%
%
%
\bibitem{ci1}
J. L. Cardy,  {\it Phase Transitions and Critical
 Phenomena\/} vol 11, edited by C. Domb and J. L. Lebowitz
  (Academic, New York, 1987). 
%
%
\bibitem{ci2}
H. W. J. Bl\"ote, J. L. Cardy and M. P. Nightingale,  
{\it Phys. Rev. Lett.\/} {\bf 56}, 742 (1986);
I. Affleck, {\it ibid.\/} {\bf 56}, 746 (1986).
%
%
\bibitem{s1c.7} F. C. Alcaraz, J. R. D. de Fel\'{\i}cio , 
R. K\"oberle and J. F.  Stilck,
 {\it Phys. Rev. B\/} {\bf 32}, 7469 (1985).
%
%
\bibitem{drugo} D. B. Balb\~ao and J. R. Drugowich de Fel\'{\i}cio, 
{\it J. Phys. A: Math. Gen.\/} {\bf 20}, L207 (1987); G. v. Gehlen, 
{\it ibid.\/} {\bf 24}, 5371 (1990).
%
%
%
\bibitem{srg}
 S. M. de Oliveira, P. M. C. de Oliveira and F. C. de S\'a Barreto, 
{\it J. Stat. Phys.\/} {\bf 78}, 1619 (1995).
%
%
\bibitem{s1.5mk} A. Bakchich, A. Bassir and A. Benyoussef, {\it Physica A\/}
{\bf 195}, 188 (1993).
%
%
\bibitem{s1.5mf} J. A. Plascak, J. G. Moreira and F. C. S\'a Barreto,
 {\it Phys. Lett. A\/} {\bf 173}, 360 (1993).
%
\bibitem{plandau0} J. A. Plascak and D. P. Landau, Computer Simulation Studies
in  Condensed Matter Physics XIII, D. P. Landau, S. P. Lewis and H.-B. 
Shuttler (eds)  (Springer, Berlin, Heidelberg 2000).
%
\bibitem{plandau} J. A. Plascak and D. P. Landau, Phys. Rev. E {\bf 67},
 R015103 (2003).
%
%
\bibitem{bw} R. J. Baxter  and F. Y. Wu, 
{\it Phys. Rev. Lett.\/} {\bf 31}, 1294 (1973); 
R. J. Baxter  and F. Y. Wu, {\it Aust. J. Phys.\/} 
{\bf 27}, 357 (1974); 
R. J. Baxter,  {\it Aust. J. Phys.\/} {\bf 27}, 369 (1974).
%
%
\bibitem{ax1} F. C. Alcaraz  and J. C. Xavier, 
{\it J. Phys. A: Math. Gen.\/} {\bf 30}, L203 (1997).
%
%
\bibitem{ax2} F. C. Alcaraz  and J. C. Xavier, {\it J. Phys. A: Math. Gen.\/},
{\bf 32}, 2041-2060 (1999).
%
%
\bibitem{merlini} D. Merlini and C. Gruber,  {\it J. Math. Phys.\/} {\bf 13}, 
1814 (1972); D. W. Wood and H. P. Griffiths, {\it J. Phys. C\/} {\bf 5}, L253
(1972).  
%
%
%
%
\bibitem{dwbaxter} E. Domany and E. K. Riedel,  
{\it J. App. Phys.\/} {\bf 49}, 1315
(1978); F. Y. Wu,  {\it Rev. Mod. Phys.\/} {\bf 54}, 235 (1982); R. J. Baxter,  
{\em Exactly Solved Models in Statistical Mechanics} (Academic, New York, 
1982). 
%
%
\bibitem{nis} B. Nienhuis, A. N. Berker, E. K. Riedel and M. Schick, 
{\it Phys. Rev. Lett.\/} {\bf 43}, 737 (1979). 
%
%
\bibitem{kda}
 W. Kinzel, E. Domany and A. Aharony, 
{\it J. Phys. A: Math. Gen.\/} {\bf 14}, L417 (1981).
%
%
\bibitem{indekeu1} J. O. Indekeu, A. Maritan and A. L. Stella, J. Phys. 
{\bf A}, L291 (1982).
%
%
\bibitem{indekeu2} J. O. Indekeu, A. Maritan and A. L. Stella, Phys. Rev. B
{\bf 35}, 305 (1987).
%
%
\bibitem{plarev} J. A. Plascak, W. Figueiredo and B. C. S. Grandi, Braz.
J. Phys. {\bf 29}, 3 (1999).
%
%
\bibitem{comm} 
Let $m_A$, $m_B$, and $m_C$  be the density of magnetization of the  
three sub-lattices. 
The three ferrimagnetic phases correspond to a phase where the 
magnetization $m=(m_A,m_B,m_C)$ has the following values $(i)$ $m=(+1,-1,-1)$, 
$(ii)$ $m=(-1,+1,-1)$, and $(iii)$ $m=(-1,-1,+1)$. There is also
a ferromagnetic phase at low temperature given by $m=(1,1,1)$.
%
%
\bibitem{fss}
M. N. Barber, {\it Phase Transitions and Critical
 Phenomena\/} vol 8, edited by C. Domb and J. L. Lebowitz 
  (Academic, New York, 1983).
%
%
\bibitem{tripoint}
 M. E. Fisher and N. Berker,  
{\it Phys. Rev.\/} 
 {\bf 26}, 2707 (1982); 
P. A. Rikvold, W. Kinzel, J. D. Gunton and K. Kaski  
{\it Phys. Rev. B\/} {\bf 28},  2686 (1983);
H. J. Hermann, {\it Phys. Lett. A\/} {\bf 100}, 156 (1984);
A. L. Malvezzi, {\it Braz. J. Phys.\/} {\bf 24}, 508 (1994).
%
%
\bibitem{lanc}
 G. H. Golub and C. F. van Loan, {\it Matrix Computations\/}, 3rd ed.
(The Johns Hopkins University Press, Baltimore, 1996).
%
%
\bibitem{moraes} M. L. M. Costa and J. A. Plascak, to appear in Braz. J. Phys.
%
%
\bibitem{moraes2} M. L. M. Costa and J. A. Plascak, unpublished.
%
%
\end{thebibliography}
\end{document}